\documentstyle[12pt]{article}

\newcommand{\lb}{\lbrack}
\newcommand{\rb}{\rbrack}

\newcommand{\msc}[1]{\mbox{\scriptsize #1}}
\newcommand{\dsp}{\displaystyle}

\newcommand{\br}{\mbox{{\bf R}}}
\newcommand{\bz}{\mbox{{\bf Z}}}
\newcommand{\bq}{\mbox{{\bf Q}}}

\newcommand{\bsz}{\msc{{\bf Z}}}

\newcommand{\cH}{{\cal H}}

\newcommand{\cN}{{\cal N}}
\newcommand{\cM}{{\cal M}}
\newcommand{\cF}{{\cal F}}

\newcommand{\cS}{{\cal S}}

\newcommand{\Th}[2]{\Theta_{#1,#2}}
\newcommand{\th}{{\theta}}
\newcommand{\tTh}[2]{\widetilde{\Theta}_{#1,#2}}
\newcommand{\ch}[2]{\mbox{ch}^{#1}_{#2}}

\newcommand{\tr}{\mbox{Tr}}

\newcommand{\NS}{\mbox{NS}}
\newcommand{\tNS}{\widetilde{\mbox{NS}}}
\newcommand{\R}{\mbox{R}}
\newcommand{\tR}{\widetilde{\mbox{R}}}
\newcommand{\sNS}{\msc{NS}}
\newcommand{\stNS}{\widetilde{\msc{NS}}}
\newcommand{\sR}{\msc{R}}
\newcommand{\stR}{\widetilde{\msc{R}}}

\newcommand{\ctri}{\chi^{\msc{tri}}}
\newcommand{\tctri}{\tilde{\chi}^{\msc{tri}}}

\newcommand {\eqn}[1]{(\ref{#1})}

\makeatletter
\@addtoreset{equation}{section}
\def\theequation{\thesection.\arabic{equation}}
\makeatother

\begin{document}
\begin{titlepage}

 {\baselineskip=14pt
 \rightline{
 \vbox{\hbox{hep-th/0111012}
       \hbox{UT-974}
       }}}

\vskip1.5cm

\begin{center}
{\bf \Large String Theory on $G_2$ Manifolds Based on Gepner Construction}
\end{center}

\bigskip

\vskip2cm

\begin{center}
Tohru Eguchi and Yuji Sugawara \\

\bigskip

Department of Physics, University of Tokyo\\

\bigskip

Tokyo, Japan 113-0033
\end{center}

\vskip2.5cm

\begin{abstract}
\vskip0.5cm

We study the type II string theories compactified on manifolds of $G_2$ 
holonomy 
of the type $(\mbox{Calabi-Yau 3-fold} \times S^1)/\bz_2$ where 
$CY_3$ sectors realized by the Gepner models.
We construct modular invariant partition functions for $G_2$ manifold 
for arbitrary Gepner models of the Calabi-Yau sector.
We note that
the conformal blocks contain the 
tricritical Ising model
and find extra massless states in
the twisted sectors of the theory when all the levels $k_i$ 
of minimal models in Gepner constructions are even.

\end{abstract}

\end{titlepage}


\section{Introduction}

Recently a number of papers have appeared investigating the dynamics of
M theory and string theory compactified on manifolds with $G_2$ holonomy 
\cite{HM}-\cite{AV2}.
M theory, when compactified on $7$-dimensional $G_2$ manifolds, provides 
${\cal N}=1$ 4-dimensional supersymmetric gauge theories which are of basic
phenomenological importance. It has also been pointed out that 
some interesting duality between gauge and gravity systems in type IIA theory 
may be interpreted as geometrical transitions when lifted to 
M-theory compactified on certain $G_2$ manifolds \cite{Ach,AMV,AW}. 
In the case of string theory compactified on 
$G_2$ manifold, on the other hand, world-sheet 
description is expected to possess some exotic
features, i.e. existence of tricritical Ising model and extended conformal
symmetry \cite{SV,OF,Gep}. Construction of Gepner type soluble models
for strings on $G_2$ manifolds has been an challenging problem.

 In a previous communication we have constructed candidate partition functions
for strings propagating on non-compact $G_2$ manifolds associated with $A-D-E$
singularities \cite{ES1}. In this paper we instead would like to 
consider the case of compact $G_2$ manifolds $(CY_3\times S^1)/{\bf Z}_2$
constructed by taking a 
Calabi-Yau 3-fold $CY_3$ times a circle $S^1$ and then by dividing 
by ${\bf Z}_2$ which acts as anti-holomorphic involution
on $CY_3$ \cite{Joyce}. 
We will use the Gepner construction for the $CY_3$ sector of  
the theory based on the tensor products of ${\cal N}=2$ minimal 
models \cite{Gepner}.
${\bf Z}_2$ orbifoldization of 
${\cal N}=2$ minimal models is somewhat non-trivial. It turns out, however, 
the representation theory and character formulas developed for the 
"twisted (orbifoldized) ${\cal N}=2$ superconformal algebra" of 
refs.\cite{ZF2,Qiu,RY2} 
provide the necessary information and we can carry out the orbifoldization 
procedure in a straightforward manner. 

In the following section we will construct a
modular invariant partition function on $G_2$ manifold 
$(CY_3\times S^1)/{\bf Z}_2$ for an arbitrary 
Gepner model describing $CY_3$. 
While the amplitude in untwisted sector contains massless states 
of $b_2+b_3=h^{1,1}+h^{2,1}+1$
chiral multiplets in 3 dimensions in agreement with geometry 
\cite{SV,HM,PP},
there appear new massless 
states (2 chiral multiplets) in the twisted sector if and only if
all the levels $k_i$ of the minimal
models of the tensor product are even. When all levels are even,
anti-holomorphic involution acts on CY manifolds without fixed points
and classically we do not expect new massless states.
Appearance of these states seem to be stringy quantum effects.
Section 3 is devoted to the case of $G_2$ manifolds constructed from
singular CY 3-folds and ALE spaces. 
Discussions and conclusions are given in section 4. 

While this paper was in preparation a preprint \cite{BB}
has appeared which discusses the construction of string theory on 
$(CY_3\times S^1)/\bz_2$ for 3 special cases of Calabi-Yau 3 folds.  
Results of \cite{BB} are completely consistent with ours.


\section{SCFT on $G_2$ Manifold $(CY_3\times S^1)/\bz_2$}

Let us construct the partition function of a string compactified
on a $G_2$ manifold $(CY_3\times S^1)/\bz_2$ with the $CY_3$ sector being
realized by a Gepner model.
Gepner model is given by a tensor product of ${\cal N}=2$ minimal models 
\begin{equation}
\left\lb \cM^{\cN=2}_{k_1}\times 
\cdots \times \cM^{\cN=2}_{k_r}\right\rb_{\msc{$U(1)$-projected}}
\equiv(k_1,\cdots,k_r) ,
\label{Gepner}
\end{equation}
where $\cM^{\cN=2}_{k}$ denotes the level $k$ $\cN=2$ minimal model
with central charge $ c = \frac{3k}{k+2}$. 
The criticality condition is given by
\begin{equation}
\sum_{i=1}^r \frac{3k_i}{k_i+2} =9.
\label{criticality CY3}
\end{equation} 
When $r=5$, the condition \eqn{criticality CY3}
becomes equivalent to the Calabi-Yau condition
for the hypersurface 
\begin{equation}
Z_1^{k_1+2}+\cdots +Z_5^{k_5+2}=0
\end{equation} 
in the weighted projective space 
$ W{\bf CP}^4(\frac{1}{k_1+2},\ldots,\frac{1}{k_5+2})$. 

The sector of the circle $S^1$ is described by 
a free boson and fermion $X$, $\psi$.
Orbifoldization along this direction is simply given by
\begin{equation}
X\,\rightarrow\, -X, ~~~\psi\,\rightarrow\, -\psi .~~~
\end{equation}

It is somewhat non-trivial to perform the orbifoldization on 
the sector of the Gepner model. 
Geometrically the $\bz_2$-action 
(we shall denote it as $\sigma$)
is an anti-holomorphic involution on $CY_3$ and 
have the properties $\sigma^*(K)=-K$, 
$\sigma^*(\Omega)= e^{i\theta}\bar{\Omega}$, 
where $K$ denotes the K\"{a}hler form and $\Omega$ is the
holomorphic 3-form.
Therefore, it is natural to assume that $\sigma$  acts on 
each sub-theory $\cM^{\cN=2}_k$ as an automorphism 
of $\cN=2$ superconformal algebra $\{T,\,J,\,G^+,\,G^-\}$;
\begin{equation}
\sigma ~:~ T\,\longrightarrow\, T,~~~J\,\longrightarrow\, -J,~~~
G^{\pm}\,\longrightarrow\, G^{\mp} ~.
\label{twist}
\end{equation}
In the computation of toroidal partition functions 
$\bz_2$-orbifoldization is enforced by $\sigma$-twisting  
along the ``space'' and ``time'' directions.
(\ref{twist}) implies, in the NS sector for instance, 
when the $\sigma$-twisting is applied 
in the spatial direction
the moding of the $G_1$ remains half-integral while that of $G_2$ is switched 
to integral values ($G^{\pm}=G_1\pm iG_2$).  
We next introduce characters of the $\cN=2$ minimal model
in various $\sigma$-twisted sectors.

\subsection{Twisted Characters in $\cN=2$ Minimal Model}

Characters of the untwisted sector of ${\cal N}=2$ theories are well-known. 
For various spin structures they are given by
\begin{equation}
\begin{array}{lll}
 \dsp \ch{k,\,(\sNS)}{l,m}(\tau,z)
& \equiv \tr_{{\cal H}^{\sNS}_{l,m}} q^{L_0-\hat{c}/8}y^{J_0}
&  = \chi^{l,0}_m (\tau,z) + \chi^{l,2}_m  (\tau,z),  \\
\dsp \ch{k,\,(\stNS)}{l,m}(\tau,z)
&  \equiv \tr_{{\cal H}^{\sNS}_{l,m}} (-1)^F q^{L_0-\hat{c}/8}y^{J_0}
&  = \chi^{l,0}_m (\tau,z) - \chi^{l,2}_m (\tau,z), \\
\dsp  \ch{k,\,(\sR)}{l,m}(\tau,z) 
& \equiv \tr_{{\cal H}^{\sR}_{l,m}} q^{L_0-\hat{c}/8}y^{J_0}
&  = \chi^{l,1}_m(\tau,z) + \chi^{l,3}_m(\tau,z), \\
\dsp \ch{k,\,(\stR)}{l,m}(\tau,z) 
&  \equiv \tr_{{\cal H}^{\sR}_{l,m}}(-1)^F q^{L_0-\hat{c}/8}y^{J_0}
&  = \chi^{l,1}_m (\tau,z) - \chi^{l,3}_m(\tau,z).
\end{array}
\end{equation}
Here we set 
\begin{equation}
\chi_m^{l,s}(\tau,z)=\sum_{r\in \bsz_k}c^{(k)}_{l,m-s+4r}(\tau)
\Th{2m+(k+2)(-s+4r)}{2k(k+2)}(\tau,z/(k+2)),
\end{equation}
and $c^{(k)}_{l,m}$ denotes the string function associated to 
the affine $SU(2)$ algebra at level $k$.


Let us now consider sectors with $\sigma$-twisting.
We denote the twisted characters as $\ch{k\,(I)}{l\,(S,T)}(\tau)$ where
$I$ runs over the spin structures $\NS, \tNS, \R, \tR$, and 
$S,T=\pm$ describes the spatial 
and temporal boundary conditions of the $\sigma$-twist. 
Since the twisting 
$\sigma ~:~J\,\rightarrow\,-J$ leaves only the states with vanishing
$U(1)$-charge, it is obvious that the twisted characters 
are labeled only by  the ``$l$-index''. 
Recall that the usual twisting by $(-1)^F$ insertion acts as
\begin{equation}
(-1)^F~:~T\,\longrightarrow\, T,~~~J\,\longrightarrow\, J,~~~
G^{\pm}\,\longrightarrow\, -G^{\pm} .
\label{GSO}
\end{equation}
Thus under the combined twist $\sigma\cdot (-1)^F 
(\equiv  (-1)^F \cdot \sigma$)
we have
\begin{equation}
\sigma\cdot(-1)^F~:~T\,\longrightarrow\, T,~~~J\,\longrightarrow\, -J,~~~
G^{\pm}\,\longrightarrow\, -G^{\mp} .
\label{twist 2}
\end{equation}
(\ref{twist 2}) differs from (\ref{twist}) only in the exchange of
$G_1$ and $G_2$ and thus leads to the same character formulas.
These facts imply the following relations among the twisted
characters;
\begin{eqnarray}
\ch{k\,(\sNS)}{l\,(+,-)}(\tau) = \ch{k\,(\stNS)}{l\,(+,-)}(\tau),~~~
\ch{k\,(\sNS)}{l\,(-,+)}(\tau) = \ch{k\,(\sR)}{l\,(-,+)}(\tau), ~~~
\ch{k\,(\stNS)}{l\,(-,-)}(\tau) = \ch{k\,(\sR)}{l\,(-,-)}(\tau),
\label{twisted character 1} \\
\ch{k\,(\sR)}{l\,(+,-)}(\tau) = \ch{k\,(\stR)}{l\,(+,-)}(\tau),~~~
\ch{k\,(\stNS)}{l\,(-,+)}(\tau) = \ch{k\,(\stR)}{l\,(-,+)}(\tau), ~~~
\ch{k\,(\sNS)}{l\,(-,-)}(\tau) = \ch{k\,(\stR)}{l\,(-,-)}(\tau).
\label{twisted character 2}
\end{eqnarray}
Characters in the 2nd line above actually all vanish due to a
fermion zero mode 
and we are left with 3
independent characters which are related to each other by the 
modular transformations;
\begin{eqnarray}
\ch{k\,(\sNS)}{l\,(+,-)}(\tau)~\stackrel{S}{\longleftrightarrow}~
\ch{k\,(\sNS)}{l\,(-,+)}(\tau)~\stackrel{T}{\longleftrightarrow}~
\ch{k\,(\stNS)}{l\,(-,-)}(\tau) ,\\
\ch{k\,(\sR)}{l\,(+,-)}(\tau)~\stackrel{S}{\longleftrightarrow}~
\ch{k\,(\stNS)}{l\,(-,+)}(\tau)~\stackrel{T}{\longleftrightarrow}~
\ch{k\,(\sNS)}{l\,(-,-)}(\tau) .
\end{eqnarray}

Fortunately we can make use of the results given in \cite{ZF2,Qiu,RY2}
to calculate these character functions.
As in \cite{Qiu}, we first consider the sector $I=\NS$,
$(S,T)=(-,+)$, which is known as the
``twisted $\cN=2$ minimal model''. In this sector 
$J$ and $G_1$ have half-integer modes and $G_2$ has 
integer modes. Making use of the well-known decomposition \cite{ZF2,Qiu2}
\begin{equation}
\cM^{\cN=2}_k \cong \frac{[\mbox{$\bz_k$-parafermion theory}] \, \times \,
    U(1)}{\bz_k} ~,
\end{equation}
the primary fields (in $\NS$ sector) in the twisted minimal 
model are constructed as
\begin{equation}
\Phi_l(z)= \varphi_l(z)\, \sigma(z), ~~~ (l=0,1,\ldots, k),
\end{equation}
where $\varphi_l(z)$ are ``$C$-disorder fields'' \cite{ZF2}
in the $\bz_k$-parafermion  theory \cite{ZF1}
and $\sigma(z)$ is the twist field of the $U(1)$ sector.
$\Phi_l(z)$ has the conformal weight
\begin{equation}
h^t_l(\equiv h(\Phi_l)) = h(\varphi_l)+h(\sigma)= 
\frac{k-2+(k-2l)^2}{16(k+2)}+\frac{1}{16}.
\end{equation}
Since we have the field identification $\Phi_l = \Phi_{k-l}$, 
we can assume the range $ l=0,1,\ldots, \lb \frac{k}{2}\rb$.
The character of the representation associated to the field 
$\Phi_l$ has been calculated in
\cite{Dobrev} (See also \cite{Qiu,RY2});
\begin{eqnarray}
\chi^{k}_{l\,(-,+)}(\tau)(\equiv
\ch{k\,(\sNS)}{l\,(-,+)}(\tau)) &=& \frac{1}{\th_4(\tau)}
\,\left(\Th{l+1-\frac{k+2}{2}}{k+2}(\tau)-
\Th{-(l+1)-\frac{k+2}{2}}{k+2}(\tau)\right)  \label{-+}\\
&= & \frac{1}{\th_4(\tau)}\left(\Th{2(l+1)-(k+2)}{4(k+2)}(\tau)
 + \Th{2(l+1)+3(k+2)}{4(k+2)}(\tau) \right. \nonumber \\
 && \left. - \Th{-2(l+1)-(k+2)}{4(k+2)}(\tau)
 - \Th{-2(l+1)+3(k+2)}{4(k+2)}(\tau) \right). \nonumber 
\end{eqnarray}

Characters in other sectors are found from the modular transformation 
\begin{eqnarray}
\chi^{k}_{l\,(-,+)}(\tau+1)&=& 
e^{2\pi i\left(h^t_l-\frac{k}{8(k+2)}\right)}\, \chi^{k}_{l\,(-,-)}(\tau), 
\label{-+ T}\\
\chi^{k}_{l\,(-,+)}(-\frac{1}{\tau})&=&
\sum_{l'=0}^k\, S^{(k)}_{l,l'}(-1)^{l'/2}\, \chi^{k}_{l'\,(+,-)}(\tau),
\label{-+ S}
\end{eqnarray}
where we have
\begin{eqnarray}
\chi^{k}_{l\,(-,-)}(\tau)&=& 
\frac{1}{\th_3(\tau)}
\left(\Th{2(l+1)-(k+2)}{4(k+2)}(\tau)
 +(-1)^k \Th{2(l+1)+3(k+2)}{4(k+2)}(\tau) \right. \nonumber \\
 && \left. +(-1)^l \Th{-2(l+1)-(k+2)}{4(k+2)}(\tau)
 +(-1)^{k+l} \Th{-2(l+1)+3(k+2)}{4(k+2)}(\tau) \right),
\label{--}  \\
\chi^{k}_{l\,(+,-)}(\tau)& =& 
\left\{
\begin{array}{ll}
\dsp  \frac{2}{\th_2(\tau)} \left(
\Th{2(l+1)}{4(k+2)}(\tau)+(-1)^k\Th{2(l+1)+4(k+2)}{4(k+2)}(\tau)
\right)   &  ~~ (l~:~\mbox{even}) ,\\
0 & ~~(l~:~\mbox{odd}).
\end{array}
\right.
\nonumber \\
\label{+-} 
\end{eqnarray}
In (\ref{-+ S})
$ S^{(k)}_{l,l'}\equiv \sqrt{\frac{2}{k+2}}
\sin\left(\frac{(l+1)(l'+1)}{k+2}\right)$ is the 
coefficient of the S-matrix of the $SU(2)$ WZW model at level $k$.

Modular properties of $\chi^{k}_{l\,(+,-)}(\tau)$, 
$\chi^{k}_{l\,(-,-)}(\tau)$ are similarly obtained as
\begin{eqnarray}
&&\hskip-15mm
\chi^{k}_{l\,(+,-)}(\tau+1)= e^{2\pi i \left(h_l -\frac{k}{8(k+2)}\right)}\,
\chi^{k}_{l\,(+,-)}(\tau), 
\hskip2mm
\chi^{k}_{l\,(+,-)}(-\frac{1}{\tau})=\sum_{l'=0}^k\, (-1)^{l/2} 
S^{(k)}_{l,l'}\,
\chi^{k}_{l'\,(-,+)}(\tau), \label{+- S}  \\
&&\hskip-15mm \chi^{k}_{l\,(-,-)}(\tau+1)= 
e^{2\pi i \left(h^t_l -\frac{k}{8(k+2)}\right)}\,
\chi^{k}_{l\,(-,+)}(\tau)  \label{-- T}, \hskip2mm
\chi^{k}_{l\,(-,-)}(-\frac{1}{\tau})=(-i)\,
\sum_{l'=0}^k\, \widehat{S}^{(k)}_{l,l'}\,
\chi^{k}_{l'\,(-,-)}(\tau) \label{-- S}  .
\end{eqnarray}
Here $h_l\equiv \frac{l(l+2)}{4(k+2)}$ and we set $\widehat{S}^{(k)}_{l,l'}= 
e^{\frac{\pi i}{2}\left(l+l'+2-\frac{k+2}{2}\right)}\, S^{(k)}_{l,l'}$
in the last line.
In summary
\begin{eqnarray}
\chi^{k}_{l\,(-,+)}(\tau)&=&
\ch{k\,(\sNS)}{l\,(-,+)}(\tau) = \ch{k\,(\sR)}{l\,(-,+)}(\tau) , \nonumber\\
\chi^{k}_{l\,(-,-)}(\tau)&=&
\ch{k\,(\stNS)}{l\,(-,-)}(\tau) = \ch{k\,(\sR)}{l\,(-,-)}(\tau) , \nonumber\\
\chi^{k}_{l\,(+,-)}(\tau)&=&
\ch{k\,(\sNS)}{l\,(+,-)}(\tau) = \ch{k\,(\stNS)}{l\,(+,-)}(\tau) .
\end{eqnarray}
Remaining characters all vanish since they contain a free Majorana fermion with
the (P,P) boundary condition. A few remarks are in order:
\begin{enumerate}
 \item It is easy to see $\chi^{k}_{k-l\,(-,+)}(\tau)=
\chi^{k}_{l\,(-,+)}(\tau)$,
$\chi^{k}_{k-l\,(-,-)}(\tau)=\chi^{k}_{l\,(-,-)}(\tau)$ 
from the
definitions \eqn{-+}, \eqn{--}. This is consistent with the field 
identification $\Phi_{k-l}\cong \Phi_l$.
 \item $\chi^{k}_{l\,(+,-)}(\tau)$ is identified with the
trace $\tr_{\cH_{l}}\left(\sigma \, q^{L_0-\frac{c}{24}}\right)$,
where $\cH_{l}$ is the representation  space of (untwisted) $\cN=2$ 
superconformal algebra over the primary state with 
$ h=\frac{l(l+2)}{4(k+2)}$, $q=0$ in the NS sector. 
$\chi^k_{l\,(+,-)}=0$ for $l~=~\mbox{odd}$
is consistent with the fact that $q\neq 0$ states do not contribute
to the trace $\tr_{\cH_{l}}\left(\sigma \, q^{L_0-\frac{c}{24}}\right)$.
 \item We present a few examples of the twisted character formulas;\\

$k=1$ ($c=1$):
\begin{eqnarray}
&&\chi^1_{0\,(-,+)}(\tau)
= \chi^1_{1\,(-,+)}(\tau)=\sqrt{\frac{\eta}{\th_4}},~~~
\chi^1_{0\,(-,-)}(\tau)=\chi^1_{1\,(-,-)}(\tau)=\sqrt{\frac{\eta}{\th_3}},~~~
\nonumber \\
&&\chi^1_{0\,(+,-)}(\tau)=\sqrt{\frac{2\eta}{\th_2}}.
\label{ex 1}
\end{eqnarray}
\bigskip
$k=2$ ($c=3/2$):
\begin{eqnarray}
&&\hskip-15mm\chi^2_{0\,(-,+)}(\tau)=\chi^2_{2\,(-,+)}(\tau)
=\sqrt{\frac{\eta}{\th_4}}\sqrt{\frac{\th_2}{2\eta}},
\chi^2_{1\,(-,+)}(\tau)=\sqrt{\frac{\eta}{\th_4}}\sqrt{\frac{\th_3}{\eta}},
\nonumber \\
&&\hskip-15mm\chi^2_{0\,(-,-)}(\tau)=\chi^2_{2\,(-,-)}(\tau)
=\sqrt{\frac{\eta}{\th_3}}\sqrt{\frac{\th_2}{2\eta}},
\chi^2_{1\,(-,-)}(\tau)=\sqrt{\frac{\eta}{\th_3}}\sqrt{\frac{\th_4}{\eta}},
\label{ex 2}\\
&&\hskip-15mm
\chi^2_{0\,(+,-)}(\tau) = \sqrt{\frac{2\eta}{\th_2}}\cdot \frac{1}{2}\left(
\sqrt{\frac{\th_3}{\eta}}+ \sqrt{\frac{\th_4}{\eta}}\right), \hskip3mm
\chi^2_{2\,(+,-)}(\tau) = \sqrt{\frac{2\eta}{\th_2}}\cdot \frac{1}{2}\left(
\sqrt{\frac{\th_3}{\eta}}- \sqrt{\frac{\th_4}{\eta}}\right)\nonumber .
\end{eqnarray}

In deriving these formulas some theta-function identities are used.
(\ref{ex 1}) ((\ref{ex 2})) is consistent with the fact that the 
$k=1$ ($k=2$) model is
described by a free boson (a free boson and fermion).

\end{enumerate}

\subsection{Partition Function of SCFT for $G_2$ Orbifold
 $(CY_3\times S^1)/\bz_2$}

Now we are ready to discuss the construction of toroidal partition functions 
of string theory on the orbifold
$(CY_3\times S^1)/\bz_2$ where the $CY_3$ sector
is described by an arbitrary Gepner model $(k_1,k_2,\ldots,k_r)$. 
We first consider the partition function of 
the $\cN=1$ non-linear $\sigma$-model on this orbifold
(using the standard diagonal modular invariant), and then go on to
the construction of the partition function of type II string theory 
on $\br^{2,1}\times (CY_3\times S^1)/\bz_2$.

According to the standard argument of $\bz_2$-orbifold, 
partition function of $\sigma$-model has the
following form
\begin{equation}
Z_{\sigma}=\frac{1}{4}\sum_I\sum_{S,T}\,Z^{(I)}_{S,T}~,
\label{sigma-model G2}
\end{equation}
where $I$ runs over spin structures $\NS$, $\tNS$, $\R$, $\tR$ and 
$S,T=\pm$ characterize the boundary conditions for the $\sigma$-twist.
The overall factor $1/4$ comes from the ${\bf Z}_2$-orbifolding and the 
GSO projection. 

The partition function in the untwisted sector is quite simple.
If the partition function of the  Gepner model $(k_1,\cdots,k_r)$
is given by
\begin{equation}
Z_{CY_3} = \frac{1}{2}\sum_I Z_{CY_3}^{(I)},
\end{equation}
then the partition function for the orbifold is given by 
\begin{equation}
Z^{(I)}_{+,+}=Z_{CY_3}^{(I)} \cdot Z_{S^1}^{(I)}.
\label{sigma-model untwisted}
\end{equation}
Amplitudes of the $S^1$ sector $Z^{(I)}_{S^1}$ are
given by the standard expressions
\begin{eqnarray}
&& Z^{(\sNS)}_{S^1} = \left|\frac{\th_3}{\eta}\right|\,Z_{S^1}(R),~~~
Z^{(\stNS)}_{S^1} = \left|\frac{\th_4}{\eta}\right|\,Z_{S^1}(R), \nonumber\\
&&Z^{(\sR)}_{S^1} = \left|\frac{\th_2}{\eta}\right|\,Z_{S^1}(R),~~~
Z^{(\stR)}_{S^1} = \left|\frac{\th_1}{\eta}\right|\,Z_{S^1}(R) \, (\equiv 0),
\end{eqnarray}
where $Z_{S^1}(R)$ denotes the partition function of a compact free 
boson $X$ ($R$ is the radius of $S^1$). 
 We later discuss the general structure of the partition function
$Z_{CY_3}$ in the Gepner model.

Now let us turn to the twisted sectors. 
Since twisted characters include only states with 
vanishing $U(1)$-charge, the orbifoldization 
enforcing the integrality of total $U(1)$-charge acts trivially 
in these sectors. We combine the conformal blocks in, say,
the $\NS$ $(+,-)$ sector as;
\begin{equation}
\hskip-3mm Z_{+,-}^{(\sNS)} =  \sum_{l_i,\bar{l}_i=0}^{k_i} \,
\prod_{i=1}^r{\cal N}^{k_i}_{l_i,\bar{l}_i}
\ch{k_i\,(\sNS)}{l_i\,(+,-)}\ch{k_i\,(\sNS)\,*}{\bar{l}_i\,(+,-)}\, 
\left|\frac{2\eta}{\th_2}\right| \left|\frac{\th_4}{\eta}\right| 
=\sum_{l_i,\bar{l}_i=0}^{k_i} \,
\prod_{i=1}^r{\cal N}^{k_i}_{l_i,\bar{l}_i}
\chi^{k_i}_{l_i\,(+,-)}\chi^{k_i\,*}_{\bar{l}_i\,(+,-)} 
\left|\frac{\th_3\th_4^2}{\eta^3}\right|.
\end{equation}
Here ${\cal N}^{k_i}_{l_i,\bar{l}_i}$ denotes the 
coefficient matrix for the modular invariants of the 
sub-theory of level $k_i$. 
Summing over spin structures we obtain
\begin{equation}
\sum_I \,Z^{(I)}_{+,-} =  \sum_{l_i,\bar{l}_i=0}^{k_i} \,
\prod_{i=1}^r{\cal N}^{k_i}_{l_i, \bar{l}_i}
\chi^{k_i}_{l_i\,(+,-)}\chi^{k_i\,*}_{\bar{l}_i\,(+,-)}
\left(\left|\frac{\th_3\th_4^2}{\eta^3}\right| 
+\left|\frac{\th_3^2\th_4}{\eta^3}\right| \right) 
\end{equation}
As it turns out, in the case of general Gepner model
describing CY 3-fold we have to be careful in choosing the coefficient 
matrices 
${\cal N}^{\{k_i\}}_{\{l_i\},\{\bar{l}_i\}}$ in order to 
ensure a suitable projection onto $\bz_2$ invariant states. 
When not all levels $k_i (i=1,\cdots,r)$ are even, we can
use the diagonal invariant for all sub-theories. 
On the other hand, in the special case with all $k_i$ even,
a particular mixture of A-type and D-type invariants has to be used
as we discuss below.
 
 Other twisted sectors are obtained from the $(+,-)$ sector by modular 
transformations. Partition functions are given by
\begin{eqnarray}
&&\sum_I \,Z^{(I)}_{-,+} = \sum_{l_i,\bar{l}_i=0}^{k_i} \,
{\cal M}^{\{k_i\}}_{\{l_i\}, \{\bar{l}_i\}}\prod_{i=1}^r
 |\chi^{k_i}_{l_i\,(-,+)}|^2 
\left(\left|\frac{\th_2\th_3^2}{\eta^3}\right| 
+\left|\frac{\th_3^2\th_2}{\eta^3}\right| \right) , 
\nonumber \\
&&\sum_I \,Z^{(I)}_{-,-} = \sum_{l_i\bar{l}_i=0}^{k_i} \,
{\cal M}^{\{k_i\}}_{\{l_i\}, \{\bar{l}_i\}}\prod_{i=1}^r 
|\chi^{k_i}_{l_i\,(-,-)}|^2 
\left(\left|\frac{\th_2\th_4^2}{\eta^3}\right| 
+\left|\frac{\th_2^2\th_4}{\eta^3}\right| \right) .
\end{eqnarray}
The matrix ${\cal M}^{\{k_i\}}_{\{l_i\},\{\bar{l}_i\}}$ is obtained from
the matrix  $\prod_i{\cal N}^{k_i}_{l_i,\bar{l}_i}$ by 
modular transformations.

Let us now consider the partition function of type II string theory 
on $\br^{2,1}\times (CY_3\times S^1)/\bz_2$. Our remaining task is to;\\
1. Fix the coefficient matrix of modular invariant.\\
2. Incorporate the contribution from the space-time $\br^{2,1}$ 
(we only consider the transversal degrees of freedom).\\
3. Take account of the GSO projection as the type II theory. 
Namely, we sum over the spin structures of left and right
movers independently, while the $\sigma$-twist acts in a diagonal 
manner.   
Due to 2 and 3, the partition function should have the following
form;
\begin{equation}
Z_{\msc{string}} = \frac{1}{4\cdot 2}\frac{1}{\sqrt{\tau_2}|\eta|^2}\, 
\sum_{I_L,I_R}\sum_{S,T}\,Z_{S,T}^{(I_L,I_R)},\hskip3mm \tau_2=\mbox{Im}\tau.
\label{string G2}
\end{equation}
where we factored out the contribution from the transverse boson of
$\br^{2,1}$ 
while that of the fermion is incorporated 
in $Z_{S,T}^{(I_L,I_R)}$ to take account of the GSO projection.
The overall factor $1/4$ is due to GSO projection while an additional $1/2$
is due to ${\bf Z}_2$-orbifolding.

Let us now introduce some formulas obtained 
in \cite{EOTY} which are convenient for the 
discussion of the general structure of 
Gepner models (see Appendix B). Contributions of the
tensor product of minimal models are organized into orbits ${\cal F}_i$
generated by the spectral flow
\begin{equation}
\cF_i(\tau)\equiv\frac{\th_3}{\eta}\,\NS_i(\tau)-
\frac{\th_4}{\eta}\,\tNS_i(\tau)-
\frac{\th_2}{\eta}\,\R_i(\tau) 
-\frac{\th_1}{\eta}\,\tR_i(\tau)
\end{equation}
where $\NS_i(\tau)$, $\tNS_i(\tau)$,  $\R_i(\tau)$, $\tR_i(\tau)$
are the conformal blocks of the $CY_3$ sector defined by    
\begin{equation}
Z_{CY_3}=\frac{1}{2}\sum_i\,D_i\left(|\NS_i(\tau)|^2+|\tNS_i(\tau)|^2
+|\R_i(\tau)|^2+|\tR_i(\tau)|^2\right) ,
\end{equation}
$D_i$ are non-negative integers with the properties 
\begin{equation}
D_i{\cal S}_{ij}=D_j{\cal S}_{ij} \hskip5mm (\mbox{no sum on i, j})
\end{equation}
where ${\cal S}_{ij}$ is the 
$S$-transformation matrix of 
the conformal blocks ${\cal F}_i$. Blocks ${\cal F}_i$ actually all 
vanish identically ${\cal F}_i\equiv 0$ 
due to some theta-function identity reflecting the space-time SUSY in 
Calabi-Yau compactification.
After a little algebra, we obtain string theory amplitude 
in the untwisted sector 
\begin{eqnarray}
\sum_{I_L,I_R}\,Z^{(I_L,I_R)}_{+,+} &=& Z_{S^1}(R)\,
\sum_i\, D_i\left|\cF_i(\tau)\right|^2.  
\end{eqnarray}

Now we consider the twisted sector $(+,-)$ and discuss a suitable projection 
onto $\bz_2$ invariant states when combined with the untwisted sector.
Under the action of $\bz_2$ symmetry $U(1)$ charge flips sign and thus in the 
twisted sector we should consider only neutral states. Let us consider a 
state, for instance,
$\dsp \ch{k,\,(\sNS)}{l,m=0}$ in the NS sector of a sub-theory.
In the orbit of this state generated by the spectral flow, 
there appears another neutral state $\dsp \ch{k,\,(\sNS)}{k-l,m=0}$
if $k$ is even. These two representations of spin $l/2$ and $(k-l)/2$
are paired and they contribute an off-diagonal term to the partition function.
Therefore when the level $k$ is even, we have to adopt an analogue of
D-type modular 
invariant. On the other hand when $k$ is odd, 
we use the standard A-type modular invariant.  

In the case of a general tensor product of minimal models 
additional neutral states appear when all
the levels $k_i$ of sub-theories are even. In this case  
an additional neutral state in the orbit of
$\prod_i\dsp \ch{k_i,\,(\sNS)}{l_i,m=0}$ has a form
$\prod_{i\in S_1}\dsp \ch{k_i,\,(\sNS)}{k_i-l_i,m=0}$ $\prod_{j\in S_2}
\dsp \ch{k_j,\,(\sNS)}{l_j,m=0}$.  Here the two sets $S_1,S_2$ are 
defined as
\begin{eqnarray} 
&&i \in S_1 \hskip3mm \mbox{if} \hskip3mm {D\over k_i+2}=\mbox{odd},\\
&&j \in S_2 \hskip3mm \mbox{if}\hskip3mm  {D\over k_j+2}=\mbox{even}.
\end{eqnarray} 
and
\begin{equation}
D=\mbox{Least Common Multiple of } \, \{k_i+2 \,(i=1,\cdots,r)\}.
\end{equation} 
We then see that the D-type pairing has to be 
used for the sub-theories in the set
$S_1$ while
A-type invariant is used for sub-theories in $S_2$. Thus we introduce
\begin{equation}
{\cal N}^{\{k_i\}}_{\{l_i\}, \{\bar{l}_i\}}=
\prod_{i\in S_2}\delta_{l_i, \bar{l}_i}\prod_{j\in S_1}(\delta_{l_j,\bar{l}_j}
+\delta_{l_j,k_j-\bar{l}_j})
\end{equation}
Then the amplitude in the twisted sector $(+,-)$ is given by
\begin{equation}
\sum_{I_L,I_R}\,Z^{(I_L,I_R)}_{+,-} = 
\sum_{l_i,\bar{l}_i}\,
\left({\cal N}^{\{k_i\}}_{\{l_i\}, \{\bar{l}_i\}}\, \prod_{i}\,
\chi^{k_i}_{l_i\,(+,-)}\chi^{k_i\,*}_{\bar{l}_i\,(+,-)}\right)\,
\left|\sqrt{\frac{\th_3}{\eta}}\sqrt{\frac{\th_3\th_4^2}{\eta^3}}
-\sqrt{\frac{\th_4}{\eta}}\sqrt{\frac{\th_4\th_3^2}{\eta^3}}\right|^2 
\end{equation}
when all level are even. When an odd level is contained in the tensor product, 
coefficient ${\cal N}^{\{k_i\}}_{\{l_i\}, \{\bar{l}_i\}}$ is replaced by the
product of Kronecker delta's. 

Amplitudes in other twisted sectors are now 
obtained by modular transformations. When all levels are even, we obtain
\begin{eqnarray}
&&\hskip-5mm \sum_{I_L,I_R}\,Z^{(I_L,I_R)}_{-,+} =
\sum_{l_i}\,\left(1+(-1)^{\sum_{i\in S_1}l_i}\right)
\prod_i\,|\chi^{k_i}_{l_i\,(-,+)}|^2\,
\left|\sqrt{\frac{\th_3}{\eta}}\sqrt{\frac{\th_3\th_2^2}{\eta^3}}
-\sqrt{\frac{\th_2}{\eta}}\sqrt{\frac{\th_2\th_3^2}{\eta^3}}\right|^2 ,  
\label{-+string}\\
&&\hskip-5mm\sum_{I_L,I_R}\,Z^{(I_L,I_R)}_{-,-} = 
\sum_{l_i}\,\left(1+(-1)^{\sum_{i\in S_1}l_i}\right)
\prod_i\,|\chi^{k_i}_{l_i\,(-,-)}|^2\,
\left|\sqrt{\frac{\th_4}{\eta}}\sqrt{\frac{\th_4\th_2^2}{\eta^3}}
-\sqrt{\frac{\th_2}{\eta}}\sqrt{\frac{\th_2\th_4^2}{\eta^3}}\right|^2 .
\label{--string}\end{eqnarray}
In checking modular invariance of these formulas 
we have to cancel some unwanted sign factors by 
using\\
(1) $D$ is an integer divisible by a factor 4.\\
(2) The set $S_1$ is not empty and its number of elements is even.\\
(3) When a sub-theory of level $k_i$ belongs to $S_1$, $k_i\in 4{\bf Z}+2$.

These facts are easily derived by using the criticality condition 
(\ref{criticality CY3}).

When an odd level is contained in the tensor product,
a factor $(-1)^{\sum_{i\in S_1}l_i}$
is absent in the above formulas (\ref{-+string}),(\ref{--string}). 
Note that all the twisted amplitudes vanish identically,
which is consistent with the existence of SUSY in our orbifold 
construction.  

Space-time SUSY 
charges are constructed as vertex operators in the untwisted 
sector and hence are the $\bz_2$-invariant combinations of SUSY charges  
of the $CY_3$ compactification. Since the $\sigma$-twisting commutes with
$(-1)^{F_L}$, $(-1)^{F_R}$, such SUSY charges 
consistently act on the Hilbert space of twisted sectors also,
and give rise to the manifest cancelation of amplitudes
in twisted sectors. Thus 
our string vacuum possesses the space-time SUSY charges 
which are half as many as those of Calabi-Yau compactification
$ \frac{1}{2} \times 8 = 4$. This is of course the 
expected number of SUSY charges in the compactification on 
a $G_2$ manifold.

Let us next check the consistency of our results 
with the general argument by Shatashvili and Vafa
\cite{SV} of string compactification on $G_2$ manifold and in particular the
existence of tricitical Ising model. 
As is shown in Appendix B, conformal blocks ${\cal F}_i$
of $CY_3$ compactification are expanded in terms of functions $g_1,g_2$
defined by
\begin{eqnarray}
g_1(\tau)&\equiv& 
\frac{\th_3}{\eta}\frac{\Th{0}{3/2}}{\eta}
-\frac{\th_4}{\eta}\frac{\tTh{0}{3/2}}{\eta}-\frac{\th_2}{\eta}
\frac{\Th{3/2}{3/2}}{\eta}~,   \\
g_2(\tau)&\equiv&
\frac{\th_3}{\eta}\frac{\Th{1}{3/2}}{\eta}
+\frac{\th_4}{\eta}\frac{\tTh{1}{3/2}}{\eta}-\frac{\th_2}{\eta}
\frac{\Th{1/2}{3/2}}{\eta}~.
\end{eqnarray}
We then use the following identities and reexpress $g_1,g_2$ in terms of
functions $F_1,F_2$ which involve tricritical Ising models  
\begin{eqnarray}
g_1(\tau)&=& \eta c^{(3)}_{0,0}(\tau) F_1(\tau)
+\eta c^{(3)}_{2,0}(\tau) F_2(\tau), \label{g f1} \\
g_2(\tau)&=& \eta c^{(3)}_{0,2} (\tau)F_1(\tau)+\eta 
c^{(3)}_{2,2} (\tau) F_2 (\tau),
\label{g f2}\\
F_1(\tau)&=& \sqrt{\frac{\th_3}{\eta}}\ctri_0
-\sqrt{\frac{\th_4}{\eta}}\tctri_0
-\sqrt{2}\sqrt{\frac{\th_2}{\eta}}\ctri_{7/16}, \\
F_2(\tau)&=& \sqrt{\frac{\th_3}{\eta}}\ctri_{1/10}
+\sqrt{\frac{\th_4}{\eta}}\tctri_{1/10}
-\sqrt{2}\sqrt{\frac{\th_2}{\eta}}\ctri_{3/80}\, .
\end{eqnarray}
Here $\ctri_h$, $\tctri_h$ denote the ($\cN=1$) 
characters of tricritical Ising model of conformal dimension $h$
and $c^{(3)}_{l,m}$ is the level 3 string function of affine $SU(2)$ algebra. 
The above relations \eqn{g f1} \eqn{g f2}
can be derived by comparing two ways of rewriting
Jacobi's identity \cite{ES1,sugiyama}
\begin{eqnarray}
0=\frac{1}{\eta^4}\left(\th_3^4-\th_4^4-\th_2^4\right)
&=& g_1(\tau) \chi_b^{SU(3)}(\tau) 
+ g_2(\tau)\left(\chi_f^{SU(3)}(\tau)+ \chi_{\bar{f}}^{SU(3)}(\tau)\right)
 \label{jacobi1}\\
&=& F_1(\tau) \chi_b^{G_2}(\tau) + F_2(\tau) \chi_f^{G_2}(\tau).
\label{jacobi2}\end{eqnarray}
Here $\chi^{SU(3)}_i$ $i=b,f,\bar{f}$
denote the level 1 $\widehat{SU}(3)$ characters of the basic,
fundamental and anti-fundamental representations, and 
$\chi^{G_2}_i$ $i=b,f$ denotes the level 1 $\widehat{G}_2$
character of the basic and fundamental representations.
We also remark that 
\begin{equation}
(G_2)_1/SU(3)_1 \cong SU(2)_3/U(1)_3 \cong \bz_3\mbox{-Parafermion},
\end{equation}
as pointed out in \cite{sugiyama}.

Above formulas \eqn{jacobi1}, \eqn{jacobi2}
show that in CY compactification of $SU(3)$ holonomy
branching functions $g_1,g_2$ should necessarily 
appear in the CFT description while
functions $F_1,F_2$ should appear in compactification on $G_2$
manifold. In fact $F_1,F_2$ contain tricritical Ising model as claimed by
Shatashvili and Vafa. All these functions $g_i,F_i$ vanish due to the Jacobi 
identity.

Let us next look at the massless spectrum contained in our amplitudes.
It is easy to identify the massless states in the untwisted sector;
they are nothing but the $\bz_2$-invariant combinations of 
the massless states in the string theory on 
$\br^{2,1}\times S^1 \times CY_3$. It is straightforward to
count these states and it is known that in addition to the gravity 
multiplet there exist $b_2+b_3=h^{1,1}+h^{2,1}+1$ massless chiral fields
where $h^{1,1},h^{2,1}$ are the Hodge numbers of Calabi-Yau 3-fold 
\cite{SV,HM,PP}.

The extra massless states originating
from the twisted sectors are somewhat non-trivial. 
We first recall the formula for conformal weights of primary fields in 
the twisted $\cN=2$ minimal model 
\begin{equation}
h^t_l(\equiv h(\Phi_l)) = \frac{k-2+(k-2l)^2}{16(k+2)} + \frac{1}{16}.
\end{equation}  
Thus we find
\begin{equation}
h^t_l - \frac{k}{8(k+2)} = 
\frac{1}{4(k+2)}\left(l+1-\frac{k+2}{2}\right)^2 \geq 0 ~.
\label{inequality htl}
\end{equation}
Therefore, when $k$ is even, the inequality \eqn{inequality htl} 
is saturated at $l= k/2$, while when $k$ is odd, there is no saturation.
This leads to the following rules on the existence of 
extra massless states;\\
1. In the case when at least one of $k_i$ is odd in the tensor product of
minimal models  $(k_1,\ldots,k_r)$, 
there are no massless states in the twisted sector.\\
2. In the case when all the levels $k_i$ are even, 
we have $2\times 2=4$ massless bosonic states in the twisted sector as
is read off from the above partition function 
(a factor 2 corresponds to the choice of NS-NS,
R-R sectors). These form 2 massless chiral multiplets.

\section{$G_2$ Manifolds associated to Singular Calabi-Yau and $K_3$ 
Spaces}

Finally we  construct the partition functions of $G_2$-orbifolds
based on the Calabi-Yau spaces with isolated singularities.
We here focus on the case of $A_{k+1}$-type singularity.
The conformal system describing such Calabi-Yau space 
is given by the Gepner model like construction;
\begin{equation}
 \left\lb \cM_k^{\cN=2} \times (\mbox{$\cN=2$ Liouville}) 
\right\rb_{\msc{$U(1)$-projected}}.
\end{equation}
In addition to the minimal model we have an ${\cal N}=2$ Liouville system
which is necessary to describe the non-compact geometry of target manifold. 
The partition function of this system is studied in detail in 
\cite{ES2}.

The $\sigma$-twist acts on the minimal sector $\cM_k^{\cN=2}$
in the same way as before, and acts on the $\cN=2$ Liouville fields
$\phi$, $Y$, $\psi^{\phi}$, $\psi^Y$
as follows;
\begin{equation}
\sigma~:~ \phi~\rightarrow~ \phi,~~~Y~\rightarrow~-Y,~~~
\psi^{\phi}~\rightarrow~\psi^{\phi},~~~ 
\psi^{Y}~\rightarrow~-\psi^{Y}~~~
\end{equation}
These fields contribute extra theta functions to  the twisted sectors.
For example, in the $(\NS,~ (+,-))$-sector,
we obtain an extra factor
\begin{equation}
\left|\frac{2\eta}{\th_2}\right| \cdot 
\left|\frac{\th_3\th_4}{\eta^2}\right|
= \left|\frac{\th_3^2\th_4^2}{\eta^4}\right|.
\end{equation}

String theory partition functions are then given by
\begin{eqnarray}
Z_{\msc{string}}=
\frac{1}{\tau_2|\eta|^4}\frac{1}{8}\,\sum_{I_L,I_R}\,\sum_{S,T}\,
Z_{S,T}^{(I_L,I_R)}  
= \frac{1}{\tau_2|\eta|^4}\frac{1}{8}\,
\left(Z^{\msc{u}}_{\msc{string}}
+ Z^{\msc{t}}_{\msc{string}}\right).
\end{eqnarray}
The factor $ \frac{1}{\tau_2|\eta|^4}$ comes from the contribution 
of the Liouville field $\phi$ as well as the transverse boson.
The conformal blocks in the untwisted sector
has  essentially the same form as given in \cite{ES2},
\begin{eqnarray}
Z^{\msc{u}}_{\msc{string}}&=& \sum_{l,r}\,Z_{S_1}(R)\, 
|\cF^{(k)}_{l,r}|^2 ,\\
\cF^{(k)}_{l,r}(\tau) &=& \frac{1}{2}\, \sum_{m\in\bsz_{4(k+2)}}\,
 \frac{1}{\eta}\Th{(k+4)m+(k+2)r}{2(k+2)(k+4)}(\tau)    \\
&&  \hspace{1cm}  \times
 \left\{\left(\frac{\th_3}{\eta}\right)^2 \ch{k, (\sNS)}{l,m}
 -(-1)^{\frac{r+m}{2}}\left(\frac{\th_4}{\eta}\right)^2 \ch{k,(\stNS)}{l,m}
 -\left(\frac{\th_2}{\eta}\right)^2 \ch{k,(\sR)}{l,m}      \right\}.
\end{eqnarray}
The conformal blocks 
$\cF_{l,r}^{(k)}(\tau)$ identically vanish for arbitrary $l,r$ 
as discussed in \cite{ES2}
and we can further prove that these functions are expanded by
$g_1(\tau)$, $g_2(\tau)$ as in the case of usual Gepner models.

The contribution from the twisted sectors are again written in a form 
where manifest cancelation takes place
\begin{eqnarray}
Z^{\msc{t}}_{\msc{string}}&=& 
\left(\sum_{l=0}^k\,|\chi^k_{l\,(+,-)}|^2\right)\,
\left|\sqrt{\frac{\th_3}{\eta}} \sqrt{\frac{\th_3^3\th_4^4}{\eta^7}}
  - \sqrt{\frac{\th_4}{\eta}} \sqrt{\frac{\th_4^3\th_3^4}{\eta^7}}
  \right|^2  \nonumber \\
&&\hspace{1cm} + 
\left(\sum_{l=0}^k\,|\chi^k_{l\,(-,+)}|^2\right)\,
\left|\sqrt{\frac{\th_3}{\eta}} \sqrt{\frac{\th_3^3\th_2^4}{\eta^7}}
  - \sqrt{\frac{\th_2}{\eta}} \sqrt{\frac{\th_2^3\th_3^4}{\eta^7}}
  \right|^2  \nonumber \\
&&\hspace{1cm} + 
\left(\sum_{l=0}^k\,|\chi^k_{l\,(-,-)}|^2\right)\,
\left|\sqrt{\frac{\th_4}{\eta}} \sqrt{\frac{\th_4^3\th_2^4}{\eta^7}}
  - \sqrt{\frac{\th_2}{\eta}} \sqrt{\frac{\th_2^3\th_4^4}{\eta^7}}
  \right|^2.
\label{singular CY3 twisted}
\end{eqnarray}

In the case of singular manifolds theory has a mass gap due to the Liouville
background charge and there exist no massless normalizable states in the 
spectrum of the untwisted sector. However, somewhat surprisingly,  massless 
states do appear in the twisted sector. By a similar argument as before 
it is easy to see that there appear
4 massless chiral multiplets in the case of even $k$ in the twisted sector
while no massless states exist at odd $k$.


The case of orbifold $(ALE  \times T^3)/(\bz_2\times \bz_2)$
is also interesting.
We focus on the case of $A_{k+1}$-singularity and 
 assume that $T^3$ is a rectangular torus with the radii 
$R_1$, $R_2$, $R_3$ for simplicity. Let $X^1$, $X^2$, $X^3$ 
($\psi^1$, $\psi^2$, $\psi^3$) be the bosonic (fermionic) coordinates
of $T^3$.  
Discussion here becomes slightly involved, since we must consider
$\bz_2\times \bz_2$ orbifold instead of $\bz_2$ orbifolding in order to realize
the right amount of supersymmetry.

In this case the system is described by the
$SU(2)_k$ WZW model together with 
the Liouville field $\phi$, and 4 free fermions $\chi^a$ ($a=0,1,2,3$),
which form an $\cN=4$ superconformal field theory with $c=6$ \cite{OV}.

We should consider the following twists $\sigma_i$ ($i=1,2,3$)
\begin{eqnarray}
\sigma_i &&:~ \phi\,\rightarrow\,\phi,~~~
K^i\,\rightarrow\,K^i,~~~K^j\,\rightarrow\,-K^j, ~ (j\neq i),~~~
\nonumber \\
&&\chi^0\,\rightarrow\,\chi^0,~~~
\chi^i\,\rightarrow\,\chi^i,~~~\chi^j\,\rightarrow\,-\chi^j, ~ (j\neq 0, i).
\end{eqnarray} 
where $K^i$ denotes the $SU(2)_k$ currents.
The above action is identified as an automorphism in
$\cN=4$ SCA
\begin{eqnarray}
 \sigma_i~&:&~G^0 ~\rightarrow~ G^0, ~~~  G^i ~\rightarrow~ G^i, ~~~
G^j ~\rightarrow~ -G^j~~~(j\neq i),
\label{sigma G}\\
&&~J^i ~\rightarrow~J^i,~~~J^j ~\rightarrow~ -J^j~~~(j\neq i).
\label{sigma J}
\end{eqnarray}
We also assume that $\sigma_i$ act on the $T^3$ sector as
\begin{equation}
\sigma_i~:~
\left\{
\begin{array}{ll}
 X^j ~\rightarrow~ -X^j~~~(j\neq i),~~~ & X^i ~\rightarrow~X^i\\
 \psi^j ~\rightarrow~ -\psi^j~~~(j\neq i),~~~ & \psi^i ~\rightarrow~\psi^i.
\end{array}
\right. 
\end{equation}
Since only the two of $\sigma_1$, $\sigma_2$, $\sigma_3$ are independent 
($\sigma_1\sigma_2=\sigma_3$), such transformations define
a $\bz_2\times \bz_2$ orbifold of $ALE \times T^3$.

These $\sigma_i$-twists give rise to various twisted
sectors. 
At first glance, it appears hard to treat  
the sectors with different twists 
in different directions. 
However, it is easy to check that 
such twisted sectors always include
at least one free fermion with the (P,P)  boundary condition
and thus do not contribute to the amplitude.

Therefore, the remaining twisted sectors are given by 
the same twist (or absence of twist) in both space and time directions.
Only the non-trivial point  
is to see how the twists act on the $SU(2)_k$ currents $K^a$.
Fortunately, this problem has been already discussed in \cite{RY2} and 
it has been shown that the characters of ``twisted $SU(2)_k$'' are 
the same as those of twisted $\cM^{\cN=2}_k$, that is, 
\eqn{-+}, \eqn{--}, \eqn{+-} which we have studied in section 2.

Now the results of the computation are summarized as follows. 
\begin{eqnarray}
Z_{\msc{string}} &=& \frac{1}{\tau_2 |\eta|^4}\,\frac{1}{16}\,\left(
Z_{\msc{string}}^{\msc{u}}+Z_{\msc{string}}^{\msc{t}} 
\right), \\
Z_{\msc{string}}^{\msc{u}}&=& Z_{T^3}(R_1,R_2,R_3)\,
\sum_{l=0}^k\,|\cF_l(\tau)|^2 ,\\
\cF_l(\tau) &=& \chi^{(k)}_l(\tau)\left\{
\left(\frac{\th_3}{\eta}\right)^4-\left(\frac{\th_4}{\eta}\right)^4-
\left(\frac{\th_2}{\eta}\right)^4 
\right\},  \\
Z_{\msc{string}}^{\msc{t}}&=& \left(\sum_{j=1}^3\,Z_{S^1}(R_j)\right)\,
\left\{
\left(\sum_{l=0}^k\,|\chi^k_{l\,(+,-)}|^2 \right)\, \left|
\sqrt{\frac{\th_3}{\eta}} \sqrt{\frac{\th_3^3\th_4^4}{\eta^7}}
-\sqrt{\frac{\th_4}{\eta}} \sqrt{\frac{\th_4^3\th_3^4}{\eta^7}}
\right|^2 \right.  \nonumber \\
&& \hspace{1cm} 
+ \left(\sum_{l=0}^k\,|\chi^k_{l\,(-,+)}|^2 \right)\, \left|
\sqrt{\frac{\th_3}{\eta}} \sqrt{\frac{\th_3^3\th_2^4}{\eta^7}}
-\sqrt{\frac{\th_2}{\eta}} \sqrt{\frac{\th_2^3\th_3^4}{\eta^7}}
\right|^2 \nonumber \\
&& \hspace{1cm} \left. 
+ \left(\sum_{l=0}^k\,|\chi^k_{l\,(-,-)}|^2 \right)\, \left|
\sqrt{\frac{\th_4}{\eta}} \sqrt{\frac{\th_4^3\th_2^4}{\eta^7}}
-\sqrt{\frac{\th_2}{\eta}} \sqrt{\frac{\th_2^3\th_4^4}{\eta^7}}
\right|^2 \right\} \, .
\label{ALE twisted}
\end{eqnarray}

Again all the normalizable states in the untwisted sector are massive.
In the twisted sectors we find 6 massless chiral multiplets 
in the case of even $k$ while no massless states exist 
when $k$ is odd.

\section{Discussions and Conclusions}
 
In this paper we have constructed partition functions of type II string 
theory compactified on general 
$G_2$ manifolds of the type $(CY_3\times S^1)/{\bf Z}_2$
by making use of Gepner construction for the CY sector. It turned out 
there appear extra massless states in the twisted sector if and only if all the
levels $k_i$ of the minimal sub-theories are even. This seems somewhat 
problematic since in these cases the corresponding hypersurface
\begin{equation}
\sum_i z_i^{k_i+2}=0
\end{equation}
do not have fixed points under anti-holomorphic involution and we do not
expect new massless states to emerge. In \cite{BB} a possible resolution of
this problem is suggested based on the behavior of the NS $B$ field
taking discrete values in $G_2$ manifolds. Related problem exists in the
$G_2$ manifolds with $A-D-E$ singularities fibered over $S^3$ which
feature in gauge/gravity duality \cite{Ach,AMV}. Since the moduli
of the metric preserving $G_2$ structure is given by $b_3$ there exists no 
smooth resolution of $A-D-E$ singularities. These are interesting issues
which require further study.

\bigskip

{\it Note added}:

\vspace{.2cm}

Very recently a new preprint has appeared \cite{RW}
where it is also pointed out that there exist  extra massless states 
in $(CY_3\times S^1)/{\bf Z}_2$ theory when all
levels are even in the Gepner construction.

Research of T.E. and Y.S. are supported in part by the grant 
Special Priority Area no.707 "Supersymmetry and Unified Theory of Elementary 
Particles".

\vskip2cm

\section*{Appendix A: ~ Conventions for Theta Functions}

\setcounter{equation}{0}
\def\theequation{A.\arabic{equation}}

 Set $q:= e^{2\pi i \tau}$, $y:=e^{2\pi i z}$;
 \begin{equation}
 \begin{array}{l}
 \dsp \th_1(\tau,z)=i\sum_{n=-\infty}^{\infty}(-1)^n q^{(n-1/2)^2/2} y^{n-1/2}
  \equiv 2 \sin(\pi z)q^{1/8}\prod_{m=1}^{\infty}
    (1-q^m)(1-yq^m)(1-y^{-1}q^m), \\
 \dsp \th_2(\tau,z)=\sum_{n=-\infty}^{\infty} q^{(n-1/2)^2/2} y^{n-1/2}
  \equiv 2 \cos(\pi z)q^{1/8}\prod_{m=1}^{\infty}
    (1-q^m)(1+yq^m)(1+y^{-1}q^m), \\
 \dsp \th_3(\tau,z)=\sum_{n=-\infty}^{\infty} q^{n^2/2} y^{n}
  \equiv \prod_{m=1}^{\infty}
    (1-q^m)(1+yq^{m-1/2})(1+y^{-1}q^{m-1/2}), \\
 \dsp \th_4(\tau,z)=\sum_{n=-\infty}^{\infty}(-1)^n q^{n^2/2} y^{n}
  \equiv \prod_{m=1}^{\infty}
    (1-q^m)(1-yq^{m-1/2})(1-y^{-1}q^{m-1/2}) .
 \end{array}
 \end{equation}
 \begin{eqnarray}
 \Th{m}{k}(\tau,z)&=&\sum_{n=-\infty}^{\infty}
 q^{k(n+\frac{m}{2k})^2}y^{k(n+\frac{m}{2k})} ,\\
 \tTh{m}{k}(\tau,z)&=&\sum_{n=-\infty}^{\infty} (-1)^n
 q^{k(n+\frac{m}{2k})^2}y^{k(n+\frac{m}{2k})}.
 \end{eqnarray}
 We use abbreviations; $\th_i \equiv \th_i(\tau, 0)$
 ($\th_1\equiv 0$), $\Th{m}{k}(\tau) \equiv \Th{m}{k}(\tau,0)$,
 $\tTh{m}{k}(\tau) \equiv \tTh{m}{k}(\tau,0)$.
 We also set
 \begin{equation}
 \eta(\tau)=q^{1/24}\prod_{n=1}^{\infty}(1-q^n).
 \end{equation}
The character of $\widehat{SU}(2)_k$ with spin $l/2$ ($0\leq l \leq k$) is 
given by
\begin{equation}
\chi^{(k)}_l(\tau, z) =\frac{\Th{l+1}{k+2}-\Th{-l-1}{k+2}}
                        {\Th{1}{2}-\Th{-1}{2}},
\end{equation}
and the string function $c^{(k)}_{l,m}(\tau)$ is defined by
\begin{equation}
\chi^{(k)}_l(\tau, z)= \sum_{m\in \bsz_{2k}}c^{(k)}_{l,m}(\tau)
\Th{m}{k}(\tau,z).
\end{equation}

\section*{Appendix B: ~ Review on the Partition Functions of SCFT for 
$CY_3$ $\sigma$-Model}

\setcounter{equation}{0}
\def\theequation{B.\arabic{equation}}

In this appendix we summarize the structure of 
the partition function of non-linear $\sigma$-model on $CY_3$ 
following \cite{EOTY}.
The partition function for $CY_3$ has the following form
\begin{equation}
Z_{CY_3}(\tau, \tilde{\tau},z,\bar{z}) = \frac{1}{2}\sum_{i=1}^{2d+d'+d''}\, 
D_i\left(|\NS_i(\tau,z)|^2
+|\tNS_i(\tau,z)|^2+|\R_i(\tau,z)|^2+|\tR_i(\tau,z)|^2\right),
\end{equation} 
where the index $i$ runs over orbits generated by the spectral flow and
$D_i$ are non-negative integers with the property
\begin{equation}
D_i \cS_{ij}=D_j\cS_{ji}, ~~~(\mbox{no sum on $i$, $j$}) .
\end{equation}
In this expression, $\cS_{ij}$ denotes the modular $S$-matrix
of the conformal blocks $\NS_i(\tau)$.
There appear various orbits in Gepner construction. 
In the NS-sector we have

\begin{enumerate}
 \item Graviton orbit: $(i=1)$
\begin{equation}
\NS_1(\tau,z)= G_1(\tau)\left(f_1(z)+f_{-1}(z)\right)
+H_1(\tau)f_0(z)~,
\end{equation}
\begin{equation}
G_1(\tau)=\sum_{n=1}^{\infty}\,g_n^{(1)}q^{n},~~~
H_1(\tau)=q^{-\frac{1}{3}}\left(1+
\sum_{n=1}^{\infty}\,h_n^{(1)}q^{n}\right)
\end{equation}

\item Massless matter orbits: $(i=2,\ldots, d-1,~
i^*\equiv i+d-1=d,\ldots, 2d-1)$
\begin{equation}
\NS_i(\tau,z)= f_1(z)+ G_i(\tau)\left(f_1(z)+f_{-1}(z)\right)
+H_i(\tau)f_0(z)~,
\end{equation}
\begin{equation}
\NS_{i^*}(\tau,z)\equiv \NS_i(\tau, -z)
= f_{-1}(z)+ G_i(\tau)\left(f_1(z)+f_{-1}(z)\right)
+H_i(\tau)f_0(z)~,
\end{equation}
\begin{equation}
G_i(\tau)=\sum_{n=1}^{\infty}\,g_n^{(i)}q^{n},~~~
H_i(\tau)=\sum_{n=1}^{\infty}\,h_n^{(i)}q^{n-\frac{1}{3}} ~.
\end{equation}
We assume $D_i=D_{i^*}$.

\item Self-conjugate matter orbits: ($j=2d,\ldots, 2d+d'-1$)
\begin{equation}
\NS_j(\tau,z)= G_j(\tau)\left(f_1(z)+f_{-1}(z)\right)
+H_j(\tau)f_0(z)~,
\end{equation}
\begin{equation}
G_j(\tau)=1+\sum_{n=1}^{\infty}\,g_n^{(j)}q^{n},~~~
H_j(\tau)=\sum_{n=1}^{\infty}\,h_n^{(j)}q^{n-\frac{1}{3}}.
\end{equation}

\item Massive orbits: ($m=2d+d',\ldots, 2d+d'+d''$)
\begin{equation}
\NS_m(\tau,z)= G_m(\tau)\left(f_1(z)+f_{-1}(z)\right)
+H_m(\tau)f_0(z)~,
\end{equation}
\begin{equation}
G_m(\tau)=\sum_{n=1}^{\infty}\,g_n^{(m)}q^{n+r_m},~~~
H_m(\tau)=\sum_{n=1}^{\infty}\,h_n^{(m)}q^{n+r_m'-\frac{1}{3}},
\end{equation}
where $r_m, r_m'\in \bq$, $r_m,r_m'>0$.
\end{enumerate} 

Here $f_Q$ denotes the level $3/2$ theta function
\begin{equation}
f_Q(z)\equiv \frac{1}{\eta(\tau)}\Th{Q}{3/2}(\tau, 2z) ~,
\end{equation}
characteristic of the $c=9$ system of ${\cal N}=2$ SCFT.

Conformal blocks in other spin structures are given by spectral flow 
\begin{equation}
\tNS_i(\tau,z)\equiv \NS_i(\tau,z+\frac{1}{2}),~~~
\R_i(\tau,z) \equiv q^{3/8}y^{3/2}\,\NS_i(\tau,z+\frac{\tau}{2}),~~~
\tR_i(\tau,z) \equiv \R_i(\tau, z+\frac{1}{2}).
\end{equation}

These blocks are compatible with the 
space-time SUSY. Namely, 
the following relation holds
\begin{equation}
\left(\frac{\th_3}{\eta}\right)\NS_i(\tau,z)-
\left(\frac{\th_4}{\eta}\right)\tNS_i(\tau,z)-
\left(\frac{\th_2}{\eta}\right)\R_i(\tau,z)-
\left(\frac{\th_1}{\eta}\right)\tR_i(\tau,z) \equiv 0,
\end{equation}

We also note that 
\begin{eqnarray}
\lim_{z\rightarrow 0} \tR_i(\tau,z) &=& I_i ~~~(\mbox{Witten index})
\nonumber \\
&=& \left\{
\begin{array}{ll}
 1 & ~~~ (i:~\mbox{massless matter orbits}) \\
 -1 & ~~~ (i:~\mbox{conjugate massless matter orbits}) \\
 0 & ~~~ (i:~\mbox{others})
\end{array}
\right.
\end{eqnarray}
Hence we obtain 
\begin{equation}
\lim_{z\rightarrow 0}\, \sum_i \, D_i |\tR_i(\tau,z)|^2
=2 \sum_{i=2}^d\, D_i = -2\chi(CY_3) ~.
\end{equation}



\end{document}